\documentclass{llncs}
\usepackage{authblk}

\usepackage{listings}
\usepackage{color}
\usepackage{amsmath}  
\usepackage{graphicx} 
\usepackage{amssymb} 
\usepackage{algpseudocode}
\usepackage{algorithm}
\usepackage{comment}
\usepackage{hyperref}
\usepackage{orcidlink}
\usepackage{amsfonts}

\usepackage{float}
\floatstyle{plaintop}
\restylefloat{table}

\title{ECDSA Cracking Methods}
\titlerunning{ECDSA Cracking Methods}

\author{
  William J. Buchanan\orcidlink{0000-0003-0809-3523}\inst{1} \and
  Jamie Gilchrist\inst{1} \and 
  Keir Finlow-Bates\orcidlink{0009-0004-8308-615X}\inst{2}
}

\institute{
Blockpass ID Lab, Edinburgh Napier University, Edinburgh.
} 

\begin{document}
\maketitle
\begin{abstract} 
The ECDSA (Elliptic Curve Digital Signature Algorithm) is used in many blockchain networks for digital signatures. This includes the Bitcoin and the Ethereum blockchains. While it has good performance levels and as strong current security, it should be handled with care. This care typically relates to the usage of the nonce value which is used to create the signature. This paper outlines the methods that can be used to break ECDSA signatures, including revealed nonces, weak nonce choice, nonce reuse, two keys and shared nonces, and fault attack.
\end{abstract}


\section{Introduction}
ECDSA has been around for over two decades and was first proposed in \cite{johnson2001elliptic}. The ECDSA method significantly improved the performance of signing messages than the RSA-based DSA method. Its usage of elliptic curve methods speeded up the whole process and supported much smaller key sizes. In 2009, Satoshi Nakamoto selected it for the  Bitcoin protocol, and it has since been adopted into Ethereum and many other blockchain methods. This paper provides a review of the most well-known methods of breaking ECDSA. 

\section{Basics of Elliptic Curve Cryptography}
One of the most basic forms of elliptic curves is:

\begin{align}
y^2=x^3+ax+b \pmod p
\end{align}

and where the elliptic curve is defined with $p$, $a$, $b$ , $g_x$, $g_y$, and $n$, and where ($g_x$,$g_y$) is a base point on our curve, and $n$ is the order of the curve.

With ECDSA, the curve used in Bitcoin and Ethereum is secp256k1, and which has the form of: 

\begin{align}
y^2=x^3+7 \pmod p
\end{align}

and where $p=2^{256}-2^{32}-977$. We can also use the NIST P256 (secp256r1) curve or the NIST-defined P521 curve. 

ECDSA signatures are non-deterministic and will change each time based on the nonce used. For the public keys, we have an $(x,y)$ point on the curve. This thus has 512 bits (for secp256k1), and where the private key is a 256-bit scalar value. 

\section{Creating the ECDSA signature}

\begin{figure*}
  \includegraphics[width=\linewidth]{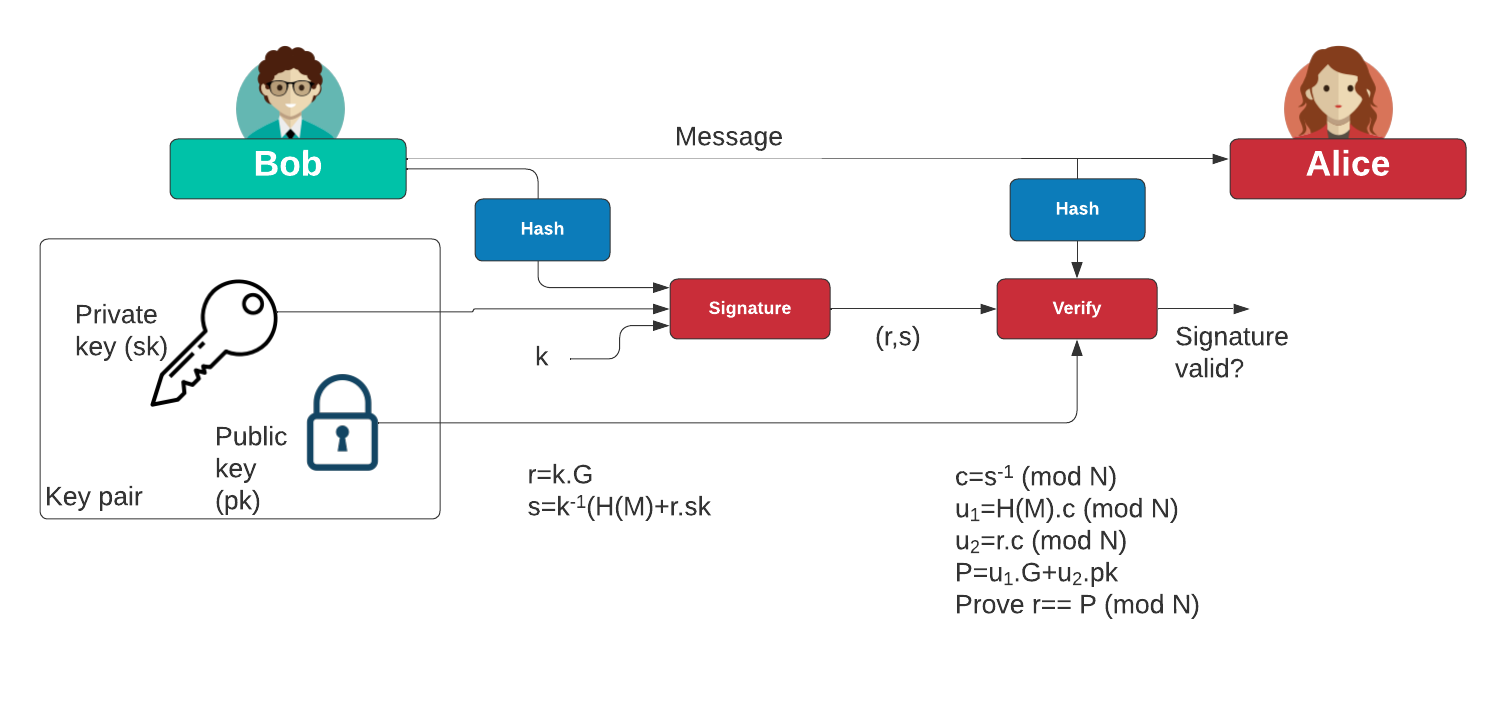}
  \caption{ECDSA signature}
  \label{fig:ecdsa}
\end{figure*}

An outline of ECDSA is shown in Figure \ref{fig:ecdsa}. With our curve, we have a generator point of $G$ and an order $n$. We start by generating a private key ($d$) and then generate the public key of:

\begin{align}
Q=d.G
\end{align}

The public key is a point on the curve, and where it is derived from adding the point $G$, $d$ times.

\subsection{Signing the message}
With a message ($m$), we aim to apply the private key and then create a signature ($r,s$). First, we create a random nonce value ($k$) and then determine the point:

\begin{align}
P=k.G
\end{align}

Next, we compute the x-axis point of this point:
\begin{align}
r=P_x \pmod n
\end{align}

This gives us the $r$ value of the signature. Next, we take the hash value of the message:
\begin{align}
e=H(m)
\end{align}

And then compute the $s$ value as:
\begin{align}
s=k^{-1}.(e + d.r) \pmod n
\end{align}

\subsection{Verifying the signature}
We can verify by taking the message ($m$, the signature ($r,s$) and the public key ($Q$):
\begin{align}
e=H(m)
\end{align}

Next, we compute:
\begin{align}
w =s^{-1} \pmod n\\
u_1 = e.w\\
u_2 = r.w
\end{align}

We then compute the point:
\begin{align}
X = u_1.G + u_2.Q 
\end{align}

And then take the x-co-ordinate of $X$:

\begin{align}
x = X_x \pmod n
\end{align}

If $x$ is equal to $r$, the signature has been verified.

\section{ECDSA attacks}

\subsection{Revealed nonce}
If the signer reveals just one nonce value by mistake, an intruder can discover the private key \cite{asecuritysite_39491}:

\begin{align}
priv= r^{-1} \times ((k \cdot s) - H(M))
\end{align}

This works because:
\begin{align}
\label{sk}
s \cdot k = H(M) + r \cdot priv
\end{align}
and so:
\begin{align}
\label{r}
r \cdot priv = s \cdot k - H(M)
\end{align}
and for $priv$:
\begin{align}
priv  = r^{-1} (s \cdot k - H(M))
\end{align}

As $r$, $s$, $H(M)$, and $k$ are known, $priv$ can therefore be calculated.

\subsection{Weak nonce choice}
A weak nonce can be broken with the Lenstra–Lenstra–Lovász (LLL) method \cite{lenstra1982factoring}, and where we crack the signature and discover the private key used to sign the message \cite{asecuritysite_79512}.

\subsection{Nonce re-use}
It is well-known that simply keeping the selected nonce secret is not enough to secure the private key \cite{brengel2018identifying}. If a nonce is used to sign a first message to produce a first signature ($s_{1}$) and is then reused to sign a second message to produce a second signature ($s_{2}$), then $s_{1}$ and $s_{2}$ will have the same $r$ value, and it is possible to derive the private key ($priv$) from the two signatures.

In ECDSA, Bob creates a random private key ($priv$) and then a public key from \cite{asecuritysite_64908}:

\begin{align}
pub = priv \times G
\end{align}

Next, to create a signature for a message $M_1$, he creates a random number ($k$) and generates the signature from the SHA-256 hash of the message, $H(M_1)$, using the private key $priv$. We denote $H(M_1)$ as $h_1$.

\begin{align}
\label{r1}
r_1 = k \cdot G\\
s_1 = k^{-1} (H(M_1) + r \cdot priv)
\end{align}

The signature is then $(r_1,s_1)$, where $r_1$ is the x-co-ordinate of the point $kG$.

Bob now signs a second message $M_2$ using the SHA-256 hash of the second message, $H(M_2)$, the same random number $k$ to produce a second signature. We denote $H(M_2)$ as $h_2$.

\begin{align}
\label{r2}
r_2 = k \cdot G\\
s_2 = k^{-1} (H(M_1) + r \cdot priv)
\end{align}

Note that $r_1$ is equal to $r_2$ (see equations \ref{r1} and \ref{r2}). In general, if two signatures generated using the same private key have the same $r$ value, then the same nonce value has been used for each signature. This provides a quick way of checking whether the nonce reused attack applies.

We can then recover the private key with:

\begin{align}
\frac{s_2 \cdot h_1 - s_1 \cdot h_2}{r(s_1-s_2)} &= \frac{h_1 \cdot h_2+r \cdot h_1 \cdot priv-h_1 \cdot h_2-r \cdot h_2 \cdot priv}{r \cdot h_1 \cdot r \cdot priv-r \cdot h_2-r \cdot priv} \\
&= \frac{r \cdot h_1 \cdot priv - r \cdot h_2 \cdot priv}{r \cdot h_1 - r \cdot h_2} \\
&= priv
\end{align}

We can also recover the nonce with:
\begin{align}
\frac{h_1-h_2}{s_1-s_2} =  \frac{h_1-h_2}{k^{-1}(h_1-h_2+priv(r-r))} = k
\end{align}

\subsection{Two keys and shared nonces}
With an ECDSA signature, we sign a message with a private key ($priv$) and prove the signature with the public key ($pub$). A random value (a nonce) is then used to randomize the signature. Each time we sign, we create a random nonce value, which will produce a different (but verifiable) signature. The private key, though, can be discovered if Alice signs four messages with two keys and two nonces \cite{brengel2018identifying}. In this case, she will sign message 1 with the first private key ($x_1$), sign message 2 with a second private key ($x_2$), sign message 3 with first private key ($x_1$) and sign message 4 with the second private key ($x_2$) The same nonce ($k_1$) is used in the signing for messages 1 and 2, and another nonce ($k_2$) is used in the signing of messages 3 and 4 \cite{asecuritysite_53129}. 

In ECDSA, Bob creates a random private key ($priv$), and then a public key from:
\begin{align}
pub= priv \cdot G
\end{align}

Next, in order to create a signature for a message of $M$, he creates a random number ($k$) and generates the signature of:
\begin{align}
r = k \cdot G\\
s = k^{-1} (H(M) + r \cdot priv)
\end{align}

The signature is then $(r,s)$ and where $r$ is the x-co-ordinate of the point $kG$. $H(M)$ is the SHA-256 hash of the message ($M$), and converted into an integer value. 
 In this case, Alice will have two key pairs and two private keys ($x_1$ and $x_2$). She will sign message 1 ($m_1$) with the first private key ($x_1$), sign message 2 ($m_2$) with a second private key ($x_2$), sign message 3 ($m_3$) with the first private key ($x_1$) and sign message 4 ($m_4$) with the second private key ($x_2$). The same nonce ($k_1$) is used in the signing of messages 1 and 2, and another nonce ($k_2$) is used in the signing of messages 3 and 4. Now let's say we have four messages ($m_1$ .. $m_4$) and have hashes of:
\begin{align}
h_1=H(m_1)\\
h_2=H(m_2)\\
h_3=H(m_3)\\
h_4=H(m_4)
\end{align}

The signatures for the messages will then be $(s_1,r_1)$, $(s_2,r_1)$, $(s_3,r_2)$, and $(s_4,r_2)$: 
\begin{align}
s_1={k_1}^{-1}(h_1 + r_1 \cdot x_1) \\
s_2={k_1}^{-1}(h_2 + r_1 \cdot x_2) \\
s_3={k_2}^{-1}(h_3 + r_2 \cdot x_1) \\
s_4={k_2}^{-1}(h_4 + r_2 \cdot x_2)
\end{align}

Using Gaussian elimination, we can also recover the private keys with:
\begin{align}
x_1=\frac{h_1 r_2 s_2 s_3 - h_2 r_2 s_1 s_3 - h_3 r_1 s_1 s_4 + h_4 r_1 s_1 s_3}{r_1 r_2 (s_1 s_4 - s_2 s_3)} 
\end{align}
and:
\begin{align}
x_2=\frac{h_1 r_2 s_2 s_4 - h_2 r_2 s_1 s_4 - h_3 r_1 s_2 s_4 + h_4 r_1 s_2 s_3}{r_1 r_2 (s_2 s_3 - s_1 s_4)} 
\end{align}

\subsection{Fault Attack}
In the case of a fault attack in ECDSA, we only require two signatures. One is produced without a fault $(r,s)$, and the other has a fault $(r_f,s_f)$. From these, we can generate the private key \cite{sullivan2022open,poddebniak2018attacking}.

In ECDSA, Bob creates a random private key ($priv$), and then a public key from \cite{asecuritysite_45188}:
\begin{align}
pub= priv \cdot G
\end{align}

Next, in order to create a signature for a message of $M$, he creates a random number ($k$) and generates the signature of:

\begin{align}
r = k \cdot G\\
s = k^{-1} (h+ r \cdot d)
\end{align}

and where $d$ is the private key and $h=H(M)$ The signature is then $(r,s)$ and where $r$ is the x-co-ordinate of the point $kG$. $h$ is the SHA-256 hash of the message ($M$), and converted into an integer value. 

Now, let's say we have two signatures. One has a fault and the other one is valid. We then have $(r,s)$ for the valid one, and $(r_f,s_f)$ for the fault. These will be:

\begin{align}
s_f=k^{-1} \cdot (h+d \cdot r_f) \\
s=k^{-1} \cdot (h+d \cdot r)
\end{align}

and where $h$ is the hash of the message. Now, if we subtract the two $s$ values, we get:
\begin{align}
s - s_f =  k^{-1} \cdot (h+d \cdot r)- k^{-1} \cdot (h+d \cdot r_f)
\end{align}
Then:
\begin{align}
s - s_f = k^{-1} \cdot (d \cdot r-d \cdot r_f) \\
k \cdot (s - s_f ) = (d \cdot r-d \cdot r_f)  \\
k = (d \cdot r - d \cdot r_f) \cdot (s-s_f )^{-1}
\end{align}

This can then be substituted in:
\begin{align}
s = k^{-1} (h + r \cdot d)
\end{align}
This gives:
\begin{align}
s= (s-s_f) \cdot {(d \cdot r-d \cdot r_f)}^{-1} \cdot (h+d \cdot r)  \\
s \cdot (d \cdot r-d \cdot r_f) = (s-s_f) \cdot (h+d \cdot r)  \\
s \cdot d \cdot r-s \cdot d \cdot r_f = s \cdot h+s \cdot d \cdot r-h \cdot s_f-d \cdot r \cdot s_f  \\
-s \cdot d \cdot r_f= s \cdot h-h \cdot s_f-d \cdot r \cdot s_f  \\
d \cdot r \cdot s_f -s \cdot d \cdot r_f= s \cdot h-h \cdot s_f \\
d \cdot (r \cdot s_f -s \cdot r_f)= h \cdot (s-s_f)
\end{align}
We can then rearrange this to derive the private key ($d$) from:
\begin{align}
d = h \cdot (s-s_f) \cdot {(s_f \cdot r-s \cdot r_f)}^{-1}
\end{align}

\section{Conclusions}
We can see that ECDSA needs to be handled carefully, especially when using the nonce value. 

\bibliographystyle{IEEEtran}
\bibliography{main}

\begin{thebibliography}{10}
\providecommand{\url}[1]{#1}
\csname url@samestyle\endcsname
\providecommand{\newblock}{\relax}
\providecommand{\bibinfo}[2]{#2}
\providecommand{\BIBentrySTDinterwordspacing}{\spaceskip=0pt\relax}
\providecommand{\BIBentryALTinterwordstretchfactor}{4}
\providecommand{\BIBentryALTinterwordspacing}{\spaceskip=\fontdimen2\font plus
\BIBentryALTinterwordstretchfactor\fontdimen3\font minus \fontdimen4\font\relax}
\providecommand{\BIBforeignlanguage}[2]{{%
\expandafter\ifx\csname l@#1\endcsname\relax
\typeout{** WARNING: IEEEtran.bst: No hyphenation pattern has been}%
\typeout{** loaded for the language `#1'. Using the pattern for}%
\typeout{** the default language instead.}%
\else
\language=\csname l@#1\endcsname
\fi
#2}}
\providecommand{\BIBdecl}{\relax}
\BIBdecl

\bibitem{johnson2001elliptic}
D.~Johnson, A.~Menezes, and S.~Vanstone, ``The elliptic curve digital signature algorithm (ecdsa),'' \emph{International journal of information security}, vol.~1, pp. 36--63, 2001.

\bibitem{asecuritysite_39491}
\BIBentryALTinterwordspacing
W.~J. Buchanan, ``Ecdsa: Revealing the private key, if nonce known (secp256k1),'' \url{https://asecuritysite.com/ecdsa/ecd3}, Asecuritysite.com, 2025, accessed: April 09, 2025. [Online]. Available: \url{https://asecuritysite.com/ecdsa/ecd3}
\BIBentrySTDinterwordspacing

\bibitem{lenstra1982factoring}
A.~K. Lenstra, H.~W. Lenstra, and L.~Lov{\'a}sz, ``Factoring polynomials with rational coefficients,'' 1982.

\bibitem{asecuritysite_79512}
\BIBentryALTinterwordspacing
W.~J. Buchanan, ``Ecdsa crack using the lenstra–lenstra–lovász (lll) method,'' \url{https://asecuritysite.com/ecdsa/ecd}, Asecuritysite.com, 2025, accessed: April 04, 2025. [Online]. Available: \url{https://asecuritysite.com/ecdsa/ecd}
\BIBentrySTDinterwordspacing

\bibitem{brengel2018identifying}
M.~Brengel and C.~Rossow, ``Identifying key leakage of bitcoin users,'' in \emph{Research in Attacks, Intrusions, and Defenses: 21st International Symposium, RAID 2018, Heraklion, Crete, Greece, September 10-12, 2018, Proceedings 21}.\hskip 1em plus 0.5em minus 0.4em\relax Springer, 2018, pp. 623--643.

\bibitem{asecuritysite_64908}
\BIBentryALTinterwordspacing
W.~J. Buchanan, ``Ecdsa: Revealing the private key, if nonce revealled (secp256k1),'' \url{https://asecuritysite.com/ecdsa/ecd5}, Asecuritysite.com, 2025, accessed: April 09, 2025. [Online]. Available: \url{https://asecuritysite.com/ecdsa/ecd5}
\BIBentrySTDinterwordspacing

\bibitem{asecuritysite_53129}
\BIBentryALTinterwordspacing
------, ``Ecdsa: Revealing the private key, from two keys and shared nonces (secp256k1),'' \url{https://asecuritysite.com/ecdsa/ecd6}, Asecuritysite.com, 2025, accessed: April 04, 2025. [Online]. Available: \url{https://asecuritysite.com/ecdsa/ecd6}
\BIBentrySTDinterwordspacing

\bibitem{sullivan2022open}
G.~A. Sullivan, J.~Sippe, N.~Heninger, and E.~Wustrow, ``Open to a fault: On the passive compromise of $\{$TLS$\}$ keys via transient errors,'' in \emph{31st USENIX Security Symposium (USENIX Security 22)}, 2022, pp. 233--250.

\bibitem{poddebniak2018attacking}
D.~Poddebniak, J.~Somorovsky, S.~Schinzel, M.~Lochter, and P.~R{\"o}sler, ``Attacking deterministic signature schemes using fault attacks,'' in \emph{2018 IEEE European Symposium on Security and Privacy (EuroS\&P)}.\hskip 1em plus 0.5em minus 0.4em\relax IEEE, 2018, pp. 338--352.

\bibitem{asecuritysite_45188}
\BIBentryALTinterwordspacing
W.~J. Buchanan, ``Cracking rsa: Fault analysis,'' \url{https://asecuritysite.com/rsa/rsa_fault}, Asecuritysite.com, 2025, accessed: April 09, 2025. [Online]. Available: \url{https://asecuritysite.com/rsa/rsa_fault}
\BIBentrySTDinterwordspacing

\end{thebibliography}

\end{document}